\documentclass[10pt,journal,final]{IEEEtran}
\IEEEoverridecommandlockouts
\usepackage{bbm}
\usepackage{amsmath}
\usepackage{amsfonts}
\usepackage[dvips]{graphicx}
\usepackage{times}
\usepackage{cite}
\usepackage{array}
\usepackage{mathrsfs}
\usepackage{multirow}
\usepackage{amssymb}

\usepackage{stfloats}
\usepackage{subfigure}
\usepackage{footnote}
\usepackage{booktabs}
\usepackage{array}
\usepackage{algorithm}
\usepackage{subeqnarray}
\usepackage{cases}
\usepackage{threeparttable}
\usepackage{xcolor}
\usepackage{hyperref}
\usepackage{epstopdf}
\usepackage{algpseudocode}
\usepackage{bm}
\usepackage{multirow}
\usepackage{adjustbox}
\usepackage{stfloats}

\def\BibTeX{{\rm B\kern-.05em{\sc i\kern-.025em b}\kern-.08em
		T\kern-.1667em\lower.7ex\hbox{E}\kern-.125emX}}

\begin{document}
\abovedisplayskip=1pt
\belowdisplayskip=1pt
\allowdisplaybreaks

\title{Deep Learning for Joint Design of Pilot, Channel Feedback, and Hybrid Beamforming in FDD Massive MIMO-OFDM Systems}

\author{Junyi Yang, Weifeng Zhu, Shu Sun, Xiaofeng Li, Xingqin Lin, and Meixia Tao
\thanks{(Corresponding authors: Meixia Tao; Shu Sun.)}
\thanks{J. Yang, S. Sun, and M. Tao are with the Department of Electronic Engineering and the Cooperative Medianet Innovation Center (CMIC), Shanghai Jiao Tong University, Shanghai, 200240, China (emails: \{yangjunyi, shusun, mxtao\}@sjtu.edu.cn)}
\thanks{W. Zhu is with the Department of Electrical and Electronic Engineering, The Hong Kong Polytechnic University, Hong Kong, SAR, China (e-mail: eee-wf.zhu@polyu.edu.hk)}
\thanks{X. Li is with Intel Corporation, Santa Clara, CA, USA (email: xiaofeng.li@intel.com)}
\thanks{X. Lin is with Nvidia Corporation,  Santa Clara, CA, USA (email: xingqinl@nvidia.com)}
}

\maketitle	
\begin{abstract}
% Fast and precise beamforming is crucial for high-quality data transmission in massive multiple-input multiple-output (MIMO) communication systems, where the hybrid analog/digital architecture is adopted to implement the precoding/combining functions. 
This letter considers the transceiver design in frequency division duplex (FDD) massive multiple-input multiple-output (MIMO) orthogonal frequency division multiplexing (OFDM) systems for high-quality data transmission. We propose a novel deep learning based framework where the procedures of pilot design, channel feedback, and hybrid beamforming are realized by carefully crafted deep neural networks. All the considered modules are jointly learned in an end-to-end manner, and a graph neural network is adopted to effectively capture interactions between beamformers based on the built graphical representation. Numerical results validate the effectiveness of our method.

\end{abstract}

\section{Introduction}

Massive multiple-input multiple-output (MIMO) is an essential technology in the fifth generation (5G) wireless systems and beyond,  owing to its remarkable capacity of beamforming towards desired directions thus significantly enhancing spectral efficiency \cite{background_mmwave1}.
Massive MIMO beamforming requires the knowledge of downlink channel state information (CSI) at the base station (BS). In time-division duplex (TDD) systems, downlink CSI can be estimated directly at the BS based on uplink transmission by channel reciprocity. In contrast, in frequency-division duplex (FDD) systems, downlink CSI acquisition typically requires the user equipment (UE) to channel estimation based on downlink pilot transmission and then feedback the CSI to the BS. Considering the signaling overhead and computational complexity, the pilot design, channel estimation, feedback mechanism, and beamforming design constitute the primary challenges for implementing massive MIMO beamforming in FDD systems.

Recently, thanks to the powerful deep learning (DL) techniques, many studies have proposed using deep neural networks to design the aforementioned modules, namely, pilot transmission, CSI estimation and feedback, and beamforming, in FDD massive MIMO systems either separately or jointly \cite{feedback_design, reviewer1_recommended1, reviewer1_recommended2,c1, reviewer1_recommended3, c3, c5, reviewer_joint, c4}. 
In particular, in works \cite{ feedback_design, reviewer1_recommended1, reviewer1_recommended2}, only one of the modules is individually optimized using DL techniques. More specifically, the work \cite{feedback_design} introduces a DNN called CsiNet for CSI feedback, while works \cite{reviewer1_recommended1, reviewer1_recommended2} utilize DNNs comprised of fully-connected and convolutional layers for beamforming design. 
To further enhance performance, works \cite{c1, reviewer1_recommended3, c3, c5} propose the DL-based method for the joint design of CSI feedback and beamforming. Therein, works \cite{c1,reviewer1_recommended3} consider beamforming design based on channels estimated by conventional methods, while works \cite{c3, c5} assume perfect CSI at the receiver. In works \cite{reviewer_joint,c4}, the pilot design, CSI feedback and beamforming are jointly optimized by DL techniques. Note that these works on joint design \cite{reviewer_joint,c4} only focus on fully digital beamforming, which requires each antenna to be connected to a dedicated RF chain, yielding potentially unaffordable hardware costs and increased power consumption \cite{background_hybrid}. 
In addition, the aforementioned works \cite{feedback_design, reviewer1_recommended1, reviewer1_recommended2,c1, reviewer1_recommended3, c3, c5, reviewer_joint, c4} only consider narrowband systems. Directly extending these methods to broadband systems can result in a substantial increase in the number of trainable parameters.

This letter considers the practical FDD broadband massive MIMO systems with orthogonal frequency division multiplexing (OFDM) modulation and hybrid analog-digital beamforming architecture. We propose a novel DL framework to realize the joint design of pilot transmission, channel feedback, and hybrid beamforming.
The main distinctions and contributions of this work in comparison to the existing literature are as follows.
First, our considered FDD massive MIMO system is more practical with OFDM-based broadband transmission using hybrid analog-digital beamforming architecture. Therein, each subchannel consists of a set of subcarriers and is associated with an individual digital beamformer, and all subchannels share a common analog beamformer. 
% This subchannel beamforming approach leverages the channel correlation between adjacent subchannels, making it more practical for real-world systems compared to methods that transmit pilots on all subchannels. 
Second, our DL-based joint design utilizes learned pilot signals and a paired vector quantized
variational auto-encoder (VQ-VAE) for channel estimation and feedback. Compared to conventional compression-reconstruction-based channel feedback methods, VQ-VAE can represent the discrete characteristics of the received signal space more accurately, thus facilitating the efficient collection and feedback of channel information. 
Third, a novel graph neural network (GNN) is proposed for hybrid beamforming and combining (HBC) design based on the channel information feedback. The proposed GNN can effectively capture the interactions between the analog and digital beamformers in broadband systems, leading to significant performance improvements.
Numerical results demonstrate that our method can consistently achieve a 16\%$\sim$21\% higher spectral efficiency comparing to existing alternatives under the same pilot length and closely approach the performance of the benchmark system with the fully digital architecture and unlimited channel feedback capacity.

\vspace{-0.1cm}

\section{System Model and Problem Formulation}
%BS和UE要用同一个N_RF么？还是要分开表示
We consider an FDD MIMO-OFDM system, where an $N_{\rm{t}}$-antenna BS with $N_{\rm{RF},\rm{t}}$ RF chains serves an $N_{\rm{r}}$-antenna UE with $N_{\rm{RF},\rm{r}}$ RF chains and $N_{\rm{s}}$ parallel data streams $\mathbf{s}[k] \in \mathbb{C}^{N_{\rm{s}} \times 1}$ over $K$ subchannels. Here, we have $N_{\rm{t}} > N_{\rm{RF},\rm{t}} \geq N_{\rm{s}}$, $N_{\rm{r}} > N_{\rm{RF},\rm{r}} \geq N_{\rm{s}}$. Let $\mathcal{K} = \{1,2,\dots,K\}$ denote the set of subchannels. 

%transceiver大概有三步，
% In the processing procedure, there is a pilot transmission stage followed by a data transmission stage. 
The whole communication procedure between the BS and UE involves three stages of pilot transmission, channel feedback and data transmission. First, in the pilot transmission stage, the BS transmits pilots to the UE over $K_{\rm{p}}$ uniformly-spaced subchannels. Let $\mathcal{K}_{\rm{p}} = \{1, M+1, \dots, (K_{\rm{p}}-1)M+1\} \in \mathcal{K}$ denote the subchannel set for pilot transmission, where $M$ represents the subchannel interval. We assume that the pilot length is $L$ and each pilot vector, denoted as $\tilde{\mathbf{s}}_l \in \mathbb{C}^{N_{\rm{RF},\rm{t}} \times 1}$ is subject to the power constraint $|| \tilde{\mathbf{s}}_l||^2_2 \le N_{\mathrm{RF},\rm{t}}$. Let $\mathcal{L} = \{1,2,\dots,L\}$ denote the set of pilot indices. Then the $l$-th received pilot signal at the $k_{\rm{p}}$-th subchannel can be expressed as
\begin{align}\label{equ:y_pilot}
\tilde{\mathbf{y}}_l[k_{\rm{p}}] = &~ \sqrt{\rho_{\rm{p}}}\tilde{\mathbf{W}}^H_{\rm{RF},\mathit{l}}\mathbf{H}[k_{\rm{p}}]\tilde{\mathbf{F}}_{\rm{RF},\mathit{l}}\tilde{\mathbf{s}}_l + \tilde{\mathbf{W}}^H_{\rm{RF},\mathit{l}}\tilde{\mathbf{n}}_l[k_{\rm{p}}], \notag \\
\quad & \quad\quad\quad\quad\quad\quad \quad \quad \quad \quad \forall   l \in \mathcal{L},k_{\rm{p}} \in \mathcal{K}_p,
\end{align}
where $\tilde{\mathbf{y}}_l[k_{\rm{p}}]$ denotes the $l$-th column of the received pilot matrix $\tilde{\mathbf{Y}}[k_{\rm{p}}]$, $\mathbf{H}[k_{\rm{p}}] \in \mathbb{C}^{ N_{\rm{r}} \times N_{\rm{t}}}$ denotes the frequency-domain channel matrix at the $k_{\rm{p}}$-th subchannel, $\rho_{\rm{p}}$ is the power for pilot transmission, $\tilde{\mathbf{n}}_l[k_{\rm{p}}] \sim \mathcal{CN}(0,\sigma^2_n\mathbf{I}_{N_{\rm{r}}})$ indicates the $k_{\rm{p}}$-th subchannel noise vectors at the $l$-th pilot transmission. $\tilde{\mathbf{F}}_{\rm{RF},\mathit{l}} \in \mathbb{C}^{ N_{\rm{t}} \times N_{\rm{RF},\rm{t}}} $ and $\tilde{\mathbf{W}}_{\rm{RF},\mathit{l}} \in \mathbb{C}^{ N_{\rm{r}} \times N_{\rm{RF},\rm{r}}}$ represent the analog beamformer and combiner in the $l$-th pilot transmission, respectively, which follow the constant modulus constraints $|[\tilde{\mathbf{F}}_{\rm{RF},\mathit{l}}]_{i,j}|^2 = \frac{1}{N_{\rm{t}}}$ and $|[\tilde{\mathbf{W}}_{\rm{RF},\mathit{l}}]_{i,j}|^2 = \frac{1}{N_{\rm{r}}}$. 

After the pilot transmission, the received pilot signals $\mathcal{R} = \{\tilde{\mathbf{Y}}[k_{\rm{p}}]\}_{k_{\rm{p}} \in \mathcal{K}_{\rm{p}}}$ are encoded into a bit stream of length $B$ by a feedback encoder, denoted as $\mathbf{q} = v_e(\mathcal{R}) \in \{0,1\}^{B \times 1}$. This bit stream is then assumed to be fed back to the BS error free. With the feedback $\mathbf{q}$, the BS recovers the received signals by a feedback decoder as $\hat{\mathcal{R}} = v_d(\mathbf{q})$, and then designs the hybrid beamformer $\mathcal{F} = \{\mathbf{F}_{\rm{RF}},\{\mathbf{F}_{\rm{BB}}[k]\}_{k \in \mathcal{K}}\}$ with $\hat{\mathcal{R}}$. Here, we denote the recovered signals as $\hat{\mathcal{R}}=\{\hat{\mathbf{Y}}[k_{\rm{p}}]\}_{k_{\rm{p}} \in \mathcal{K}_{\rm{p}}}$. For the UE, the hybrid combiner $\mathcal{W} = \{\mathbf{W}_{\rm{RF}},\{\mathbf{W}_{\rm{BB}}[k]\}_{k \in \mathcal{K}}\}$ can be designed based on its received pilot signals $\mathcal{R}$. Here, $\mathbf{F}_{\rm{RF}} \in \mathbb{C}^{N_{\rm{t}} \times N_{\rm{RF},\rm{t}}}$ and $\mathbf{W}_{\rm{RF}} \in \mathbb{C}^{N_{\rm{r}} \times N_{\rm{RF},\rm{r}}}$ represent the analog beamformer and combiner at the BS and the UE, respectively, which also follow the constant modulus constraints $|[\mathbf{F}_{\rm{RF}}]_{i,j}|^2 = \frac{1}{N_{\rm{t}}}$ and $|[\mathbf{W}_{\rm{RF}}]_{i,j}|^2 = \frac{1}{N_{\rm{r}}}$; $\mathbf{F}_{\rm{BB}}[k] \in \mathbb{C}^{N_{\rm{RF},\rm{t}} \times N_{\rm{s}}}$ and $\mathbf{W}_{\rm{BB}}[k] \in \mathbb{C}^{ N_{\rm{RF},\rm{r}} \times N_{\rm{s}}}$ represent the digital beamformer and combiner at the $k$-th subchannel, respectively. We also consider the power constraint for the hybrid analog and digital beamformers, i.e., $\sum_{k=1}^{K}||\mathbf{F}_{\rm{RF}}\mathbf{F}_{\rm{BB}}[k]||^2_F = KN_{\rm{s}}$.
Note that we propose to directly process $\mathcal{R}$ without explicitly reconstructing the channel matrix throughout the process (i.e., implicit channel estimation), which is potential to greatly reduce the signaling overhead and thus further improve the system performance. The hybrid beamformer and combiner design can be modeled as:
\begin{align}
\mathcal{F} &= f(\hat{\mathcal{R}}), \label{equ:f_BS} \\ 
\mathcal{W} &= g(\mathcal{R}). \label{equ:g_UE}
\end{align}

Finally, we adopt the fully-connected hybrid beamforming architecture at both the BS and UE for downlink data transmission, where all the subchannels share the common analog beamformer $\mathbf{F}_{\rm{RF}}$ at the BS (and the common analog combiner $\mathbf{W}_{\rm{RF}}$ at the UE), while each subchannel has its own individual digital beamformer $\mathbf{F}_{\rm{BB}}[k]$ (and digital combiner $\mathbf{W}_{\rm{BB}}[k]$) for $k \in \mathcal{K}$. Then the received signal of the $k$-th subchannel is given by
\begin{align}\label{equ:transmition}
\vspace{-0.1cm}
\mathbf{y}[k] = &~ \sqrt{\rho}\mathbf{W}^H_{\rm{BB}}[k]\mathbf{W}^H_{\rm{RF}}\mathbf{H}[k]\mathbf{F}_{\rm{RF}}\mathbf{F}_{\rm{BB}}[k]\mathbf{s}[k] \notag \\ 
\quad &+ \mathbf{W}^H_{\rm{BB}}[k]\mathbf{W}^H_{\rm{RF}}\mathbf{n}[k], ~\forall k \in \mathcal{K}.
\vspace{-0.1cm}
\end{align}
where $\rho$ and $\mathbf{n}[k] \sim \mathcal{CN}(0,\sigma^2_n\mathbf{I}_{N_{\rm{r}}}) $ denote the transmit power and noise vector, $\mathbf{s}[k] \in \mathbb{C}^{N_{\rm{s}} \times 1}$ is the information symbol which satisfies the constraint $\mathbb{E}\{\mathbf{s}[k]\mathbf{s}^H[k]\} = \mathbf{I}_{N_{\rm{s}}}$. The spectral efficiency of the system can be computed as
\begin{align}\label{equ:rate}
R = \frac{1}{K} \sum_{k \in \mathcal{K}} \log \det\left(  \mathbf{I}_{N_{\rm{s}}} +  \frac{\rho}{N_{\rm{s}}} \mathbf{\Omega}^{-1}[k] \mathbf{\Lambda}[k] \mathbf{\Lambda}^H[k] \right),
\end{align}
where $\mathbf{\Omega}[k] = \sigma^2_{n} \mathbf{W}^H_{\rm{BB}}[k] \mathbf{W}^H_{\rm{RF}} \mathbf{W}_{\rm{RF}} \mathbf{W}_{\rm{BB}}[k] \in \mathbb{C}^{ N_{\rm{s}} \times N_{\rm{s}}} $  and  $\mathbf{\Lambda}[k] = \mathbf{W}^H_{\rm{BB}}[k]\mathbf{W}^H_{\rm{RF}} \mathbf{H}[k] \mathbf{F}_{\rm{RF}}\mathbf{F}_{\rm{BB}}[k] \in \mathbb{C}^{ N_{\rm{s}} \times N_{\rm{s}}} $.

%公式要分块

Based upon the above signal processing procedure, we jointly design the pilot parameter $\mathcal{P} = \{\tilde{\mathbf{W}}_{\mathrm{RF},l},\tilde{\mathbf{F}}_{\mathrm{RF},l},\tilde{s}_l\}_{l=1}^{L}$, the feedback, and the hybrid beamforming to maximize the spectral efficiency in the massive MIMO-OFDM system. The optimization problem can be formulated as
\begin{subequations}\label{equ:problem}
\begin{align}
\mathop{\max}\limits_{ \substack{\mathcal{P}, v_e(\cdot), v_d(\cdot), \\ \mathcal{F},\mathcal{W},f(\cdot),g(\cdot)} } &  R \quad \text{in} \quad (\ref{equ:rate}) \\
\text { s.t.  } \;\;\;\quad & |[\mathbf{F}_{\rm{RF}}]_{i,j}|^2 =|[\tilde{\mathbf{F}}_{\rm{RF},\mathit{l}}]_{i,j}|^2 = \frac{1}{N_{\rm{t}}}, \forall i,j, \label{equ:cst_F} \\
\quad & |[\mathbf{W}_{\rm{RF}}]_{i,j}|^2 =|[\tilde{\mathbf{W}}_{\rm{RF},\mathit{l}}]_{i,j}|^2 = \frac{1}{N_{\rm{r}}}, \forall i,j, \label{equ:cst_W}\\
\quad & \sum_{k=1}^{K}||\mathbf{F}_{\rm{RF}}\mathbf{F}_{\rm{BB}}[k]||^2_F = KN_{\rm{s}}, \label{equ:cst_power} \\
\quad & || \tilde{\mathbf{s}}_l||^2_2 \le N_{\mathrm{RF},\rm{t}},\forall l, \label{equ:cst_pilot_digital} \\
\quad & \mathbf{q} = v_e(\mathcal{R}), \hat{\mathcal{R}} = v_d(\mathbf{q}), \\
\quad & (\ref{equ:y_pilot}), (\ref{equ:f_BS}), (\ref{equ:g_UE}).
\end{align}
\end{subequations}
The above optimization problem contains both variable optimization and function optimization. To tackle this challenging problem, we propose a DL-based approach to learn the pilot parameter $\mathcal{P}$ and to parameterize the mapping functions $v_e(\cdot)$, $v_d(\cdot)$, $f(\cdot)$ and $g(\cdot)$ by deep neural networks, whose details will be given in the next section.

%说明这个网络需要设计的是K,v以及f和g，会在下一章节着重说明结构

\section{Proposed DL-based Method}
\begin{figure}[t]
\centering
\includegraphics[width=0.45 \textwidth]{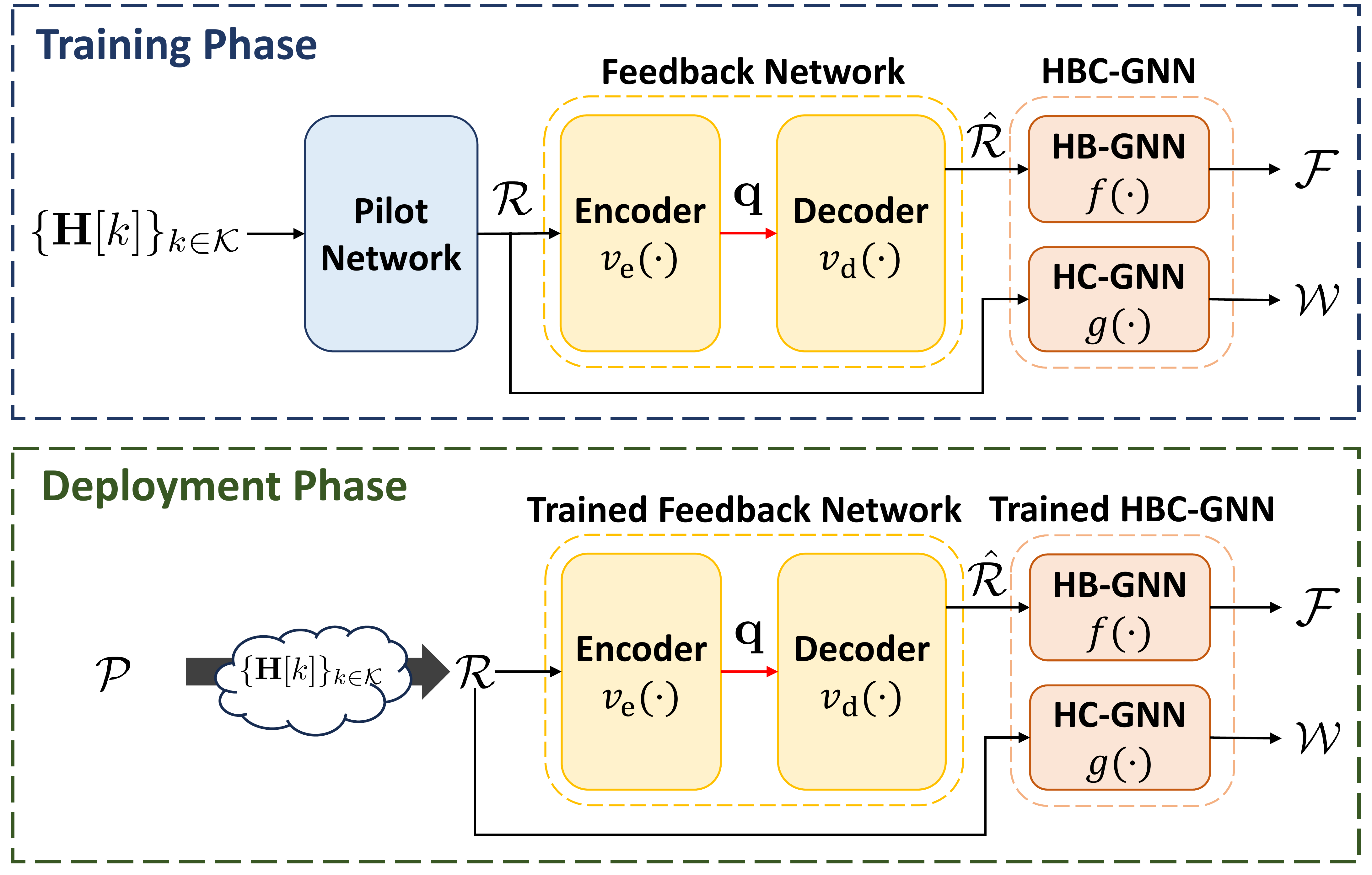}
\vspace{-0.2cm}
\caption{The diagram of the proposed DL-based method.}\label{fig:architecture}
\vspace{-0.6cm}
\end{figure}

%然后再说具体的方法
In this section, we propose a DL-based method to acquire the hybrid beamformer in the FDD massive MIMO-OFDM system. As illustrated in Fig. \ref{fig:architecture}, the proposed DL architecture consists of a pilot network (PN), a feedback
network (FN), and an HBC-GNN. 
The parameters in these DNNs are jointly optimized during the training phase before deployment.

\subsection{Pilot Network}
The PN is trained to obtain the pilot parameter $\mathcal{P}$ for channel estimation. Due to the constant-modulus constraint on each element in $\tilde{\mathbf{F}}_{\rm{RF},\mathit{l}}$ and $\tilde{\mathbf{W}}_{\rm{RF},\mathit{l}}$ (\ref{equ:cst_F})(\ref{equ:cst_W}), the PN obtains the phase shifts of them, which satisfy the equation:
\begin{subequations} \label{equ:RF}
\begin{align}
    \tilde{\mathbf{F}}_{\rm{RF},\mathit{l}} = \frac{1}{\sqrt{N_{\rm{t}}}}\left[\cos(\tilde{\pmb{\Theta}}_\mathit{l}) + \rm{j}\cdot\sin(\tilde{\pmb{\Theta}}_\mathit{l})\right], \\
    \tilde{\mathbf{W}}_{\rm{RF},\mathit{l}} =
    \frac{1}{\sqrt{N_{\rm{r}}}}\left[\cos(\tilde{\pmb{\Phi}}_\mathit{l}) + \rm{j}\cdot\sin(\tilde{\pmb{\Phi}}_\mathit{l})\right],
\end{align}
\end{subequations}
where $\pmb{\Theta}_l \in \mathbb{R}^{ N_{\rm{t}} \times N_{\rm{RF},\rm{t}}}$ and $\pmb{\Phi}_l \in \mathbb{R}^{ N_{\rm{r}} \times N_{\rm{RF},\rm{r}}}$ represent the matrices of phase shifts at the BS and the UE, respectively. Thus we consider $\{\pmb{\Theta}_l, \pmb{\Phi}_l, \tilde{\mathbf{s}}_l\}_{l=1}^{L}$ as the trainable variables and (\ref{equ:y_pilot}) can be regarded as a forward-pass computation of $\mathbf{H}[k_{\rm{p}}]$ through a two-layer network. Once the training of the PN is completed, we will directly use the trained $\mathcal{P}$ to acquire the channel information in the deployment phase.

\subsection{Feedback Network}

Based on the signal processing procedure in Section III, we can follow the auto-encoder neural network architecture to design the FN. 
In this work, we adopt the VQ-VAE neural network \cite{VQVAE} for feedback by exploiting its high-quality compression ability. 
The key idea of VQ-VAE is to train a codebook that can accurately characterize the discrete representation of the input signal space. The pilot signal $\mathcal{R}$ is first split to several vectors, then VQ-VAE utilizes the codeword closest to each vector from the trained codebook as its output. In our work, the trained codebook is pre-stored at both the UE and the BS, while the binary vector $\mathbf{q}$ in the feedback link in fact represents the indices of the selected codewords by the encoder. For illustration, we denote the loss function of VQ-VAE as $\mathcal{L}_{\mathrm{V}}$, which can be regarded as the mean squared error (MSE) between the received pilot signals and the selected codewords by the encoder.
% \footnote{Due to the page limit, the details of VQ-VAE are omitted herein and can be found in \cite{VQVAE}}

%Van Den Oord, Aaron, and Oriol Vinyals. "Neural discrete representation learning." Advances in neural information processing systems 30 (2017).

%The expression of $\mathcal{L}_{V}$ is given in [] and not specified here, due to the page limit.
%Due to the limited number of bits in the feedback link, the UE and the BS are required to adopt an auto-encoder which is denoted as $v(\cdot)$ to report the received signal of pilots. For this system, we utilize a VQ-VAE which adopts a learnable codebook (or embedding space) at the UE and the BS to realize the operation and minimize the feedback distortion caused by the quantization. 

%Specifically, the VQ-VAE firstly divides received signal $\tilde{\mathcal{M}}$ into several vectors and selects the codewords closest to them from the codebook. The indices of the selected codewords are then reported to the BS, where the same codebook is adopted to restore the information. Here, The MSE function is adopted as the loss function to fine-tune the codewords in the codebook so that they can match the vectors split by $\tilde{\mathcal{M}}$ as much as possible. Finally, we denote the feedback restored by the BS as $\tilde{\mathcal{N}} = v(\tilde{\mathcal{M}})$, which has the same data format as $\tilde{\mathcal{M}}$.

% \begin{figure*}[t]
% \begin{centering}
% \includegraphics[width=0.80 \textwidth]{GNN_architecture.pdf}
% \vspace{-0.2cm}
% \caption{xxxxxxxxxxxxxxxxxxxxxxxxxxxxxx}\label{fig:GNN}
% \end{centering}
% \vspace{-0.2cm}
% \end{figure*}

\subsection{Hybrid Beamforming and Combining Graph Neural Network}

To better exploit the interactions between the analog beamformer and digital beamformers, we utilize the GNN for hybrid beamforming design.
In the MIMO-OFDM system, there is a particular digital beamformer for each subchannel, while the analog beamformer is shared among all the subchannels. Compared with the fully-connected neural network, GNN can naturally embed the topological relations in its network architecture and thus enjoys permutation invariance and permutation equivariance of the optimization problem (\ref{equ:problem}). Here, permutation invariance means that the analog beamformer $\mathbf{F}_{\mathrm{RF}}$ is independent of the ordering of the subchannels, while permutation equivariance means that the $\{\mathbf{F}_{\mathrm{BB}}[k]\}_{k \in \mathcal{K}}$ will be permuted in the same way if the subchannels are permuted. Furthermore, the reduced model complexity and improved generalization performance also make the GNN more favorable.

\begin{figure*}[t]
  \centering
  \subfigure[]
  {\includegraphics[height=.24\textwidth]{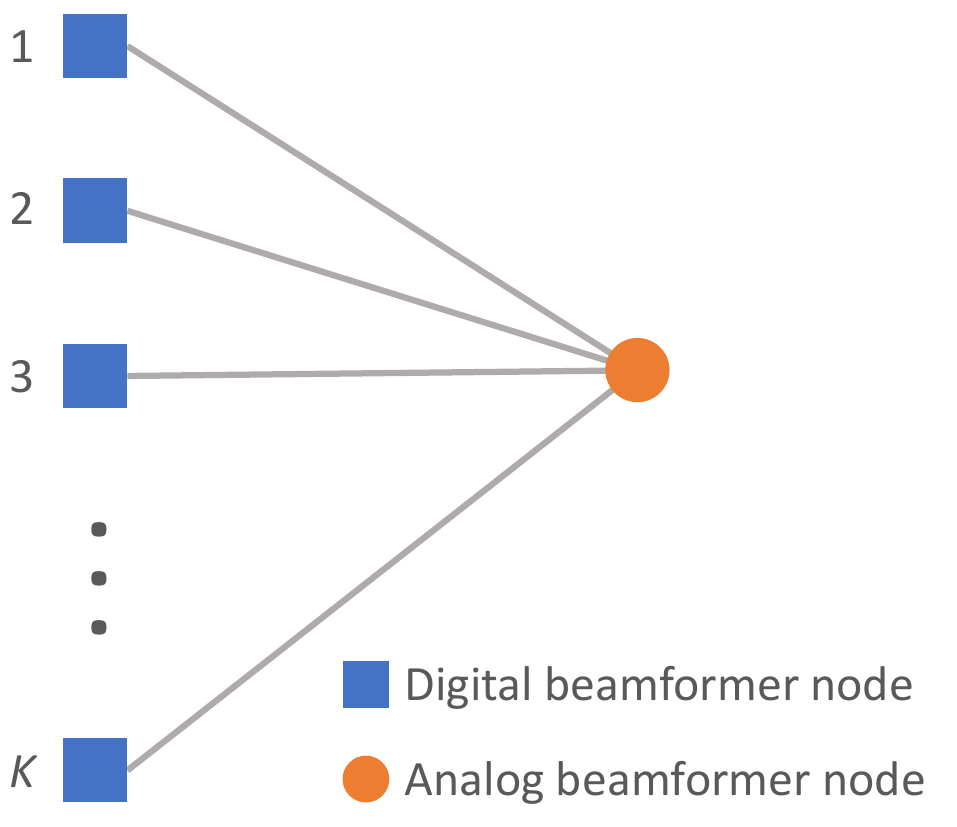}\label{fig:GR}}%\hspace{1.2cm}
  \subfigure[]
  {\includegraphics[height=.24\textwidth]{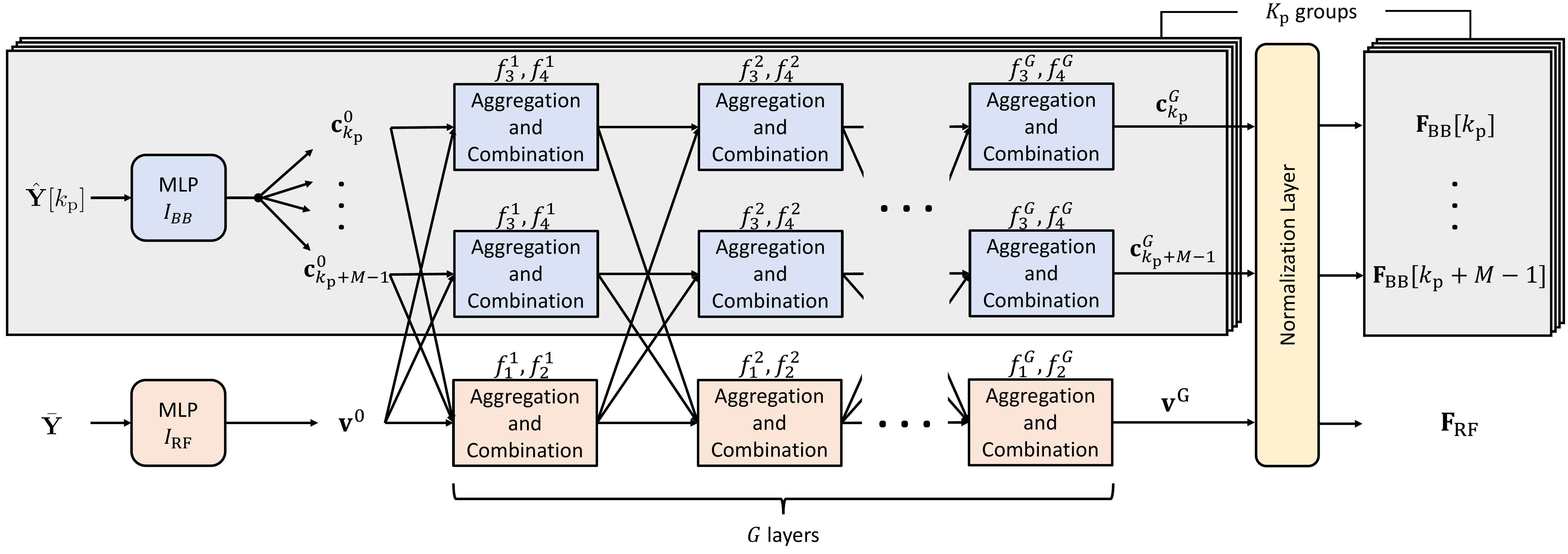}\label{fig:HB-GNN}}%\hspace{1.2cm}
  \caption{(a) Graphical representation of the hybrid beamformer at the BS; (b) The network architecture of the proposed HB-GNN.}\label{fig:GNN}
  \vspace{-0.5cm}
\end{figure*}

Before the GNN design, we build the graphical representation of $\mathbf{F}_{\mathrm{RF}}$ and $\{\mathbf{F}_{\mathrm{BB}}[k]\}_{k \in \mathcal{K}}$ at the BS. As shown in Fig. \ref{fig:GR}, the analog beamformer $\mathbf{F}_{\mathrm{RF}}$ is represented by the circular node and the digital beamformers $\{\mathbf{F}_{\mathrm{BB}}[k]\}_{k \in \mathcal{K}}$ are presented by square nodes. There is an associated state vector $\mathbf{c}_k \in \mathbb{R}^{2N_{\rm{RF},\rm{t}}N_{\rm{s}}\times 1} $ and $\mathbf{v} \in \mathbb{R}^{2N_{\rm{RF},\rm{t}}N_{\rm{t}}\times 1}$ for each digital beamformer node and analog beamformer node, respectively. These vectors will be updated layer by layer in the GNN to collect sufficient useful information, and finally yield the beamformers of their corresponding nodes. Note that the graphical representation of $\mathbf{W}_{\mathrm{RF}}$ and $\{\mathbf{W}_{\mathrm{BB}}[k]\}_{k \in \mathcal{K}}$ at the UE can be established in a similar way.

Based on the graphical representation, we propose a HBC-GNN which contains a hybrid beamforming GNN (HB-GNN) and a hybrid combining GNN (HC-GNN) to obtain $\mathcal{F}$ and $\mathcal{W}$, respectively. The HB-GNN and HC-GNN have similar network structures, thus we only provide the details of the HB-GNN below.
The architecture of the proposed HB-GNN architecture is shown in Fig.\ref{fig:HB-GNN}, which consists of the following three parts.

% Specifically, 
%这里是否需要解释因为只有部分subcarrier扫描了探测波束，所以需要先用一个mini网络初始化所有的feature vector
\subsubsection{Initialization Layer}
This layer consists of two DNNs to obtain the initialization for all nodes. Considering the correlation between adjacent subchannels,
we initialize the digital beamformer node for each subchannel based on the collected channel information from its nearby pilot-bearing subchannels. Specifically, all digital beamformer nodes are divided into $K_{\rm{p}}$ groups and the multilayer perceptron (MLP) $I_{\rm{BB}}(\cdot)$ is designed for all groups to generate initial state vectors as $\{\mathbf{c}^0_{k_{\rm{p}}},\mathbf{c}^0_{k_{\rm{p}}+1},\dots,\mathbf{c}^0_{k_{\rm{p}}+M-1}\} = I_{\rm{BB}}(\hat{\mathbf{Y}}[k_{\rm{p}}]), k_{\rm{p}} \in \mathcal{K}_{\rm{p}}$. For the analog beamformer node, the state vector is initialized by the MLP $I_{\rm{RF}}(\cdot)$ as $\mathbf{v}^0 = I_{\rm{RF}}(\bar{\mathbf{Y}})$, where $\bar{\mathbf{Y}} = \psi(\{\hat{\mathbf{Y}}[k_{\rm{p}}]\}_{k_{\rm{p}} \in \mathcal{K}_{\rm{p}}})$ and $\psi(\cdot)$ adopts the element-wise mean function due to the fact that each digital beamformer node has equal contribution to the analog beamformer node. Note that such a property is also utilized in the design of the following aggregation and combination layers.

\subsubsection{$G$ Layers of Aggregation and Combination}
In the $g$-th aggregation and combination layer, the state vector of each node is updated by combining its own state vector and the aggregation of state vectors from its neighbor nodes.

Specifically, in the $g$-th aggregation and combination layer, the state vector of the analog beamformer node is updated as
% \begin{align}
% \mathbf{v}^g = f^g_1(\mathbf{v}^{g-1}) + \psi\left(f^g_2(\mathbf{c}^{g-1}_1),\dots,f^g_2(\mathbf{c}^{g-1}_K)\right),
% \end{align}
\begin{align}
\mathbf{v}^g = f^g_1(\mathbf{v}^{g-1}) + f^g_2\left( \bar{\mathbf{c}} ^{g-1} \right) , 
\end{align}
where $\bar{\mathbf{c}} ^{g-1} = \psi(\{\mathbf{c}^{g-1}_k\}_{k \in \mathcal{K}})$, $f^g_1(\cdot)$ and $f^g_2(\cdot)$ are realized by MLPs.
% \begin{align}
% \psi(\mathbf{z}^{l}_1,\mathbf{z}^{l}_2,\dots,\mathbf{z}^{l}_M) = \frac{1}{M}\sum_{m=1}^M \mathbf{z}^{l}_m.
% \end{align}
For digital beamformer nodes, the state vector of digital beamformer node $k$ in the $g$-th layer can be given by
\begin{align}
\mathbf{c}^g_k = f^g_3(\mathbf{c}^{g-1}_k) + f^g_4(\mathbf{v}^{g-1}), ~\forall k \in \mathcal{K},
\end{align}
where $f^g_3(\cdot)$ and $f^g_4(\cdot)$ also employ MLPs and they are reused in $K_{\rm{p}}$ groups just like $I_{\rm{BB}}(\cdot)$. 
% \textcolor{red}{Here, the aggregation function is unnecessary since every digital node only has the analog node as the neighbor node.}
\subsubsection{Normalization Layer}
After $G$-layer aggregation and combination, we obtain the hybrid beamformers from the state vectors $\{\mathbf{c}^{G}_{k}\}_{k=1}^{K}$ and $\mathbf{v}^{G}$. Here, each state vector exactly consists of the real and imaginary components of its corresponding beamformer and can be represented as
\begin{align}
 \mathbf{v}^{G} &= vec(\begin{bmatrix}\Re\{\mathbf{F}_{\mathrm{RF}}\}, \Im\{\mathbf{F}_{\mathrm{RF}}\}\end{bmatrix}), \\
 \mathbf{c}^{G}_{k} &= vec(\begin{bmatrix}\Re\{\mathbf{F}_{\mathrm{BB}}[k]\}, \Im\{\mathbf{F}_{\mathrm{BB}}[k]\}\end{bmatrix}), \forall k \in \mathcal{K},
\end{align}
Then, a normalization layer is utilized to scale $\{\mathbf{F}_{\mathrm{BB}}[k]\}_{k \in \mathcal{K}}$ and each element in $\mathbf{F}_{\mathrm{RF}}$, which ensures that the constraints (\ref{equ:cst_F}), (\ref{equ:cst_W}), and (\ref{equ:cst_power}) are satisfied.

In practice, each MLP in the proposed HB-GNN and HC-GNN is modeled as only one fully-connected layer with an activation function, and the input/output dimension is determined based on the input/output vector.
%thus the computational complexity is about $O(G N^2_{\rm{RF}} (K  N^2_{\rm{s}}+ N_{\rm{s}}N+N^2 ))$, where $N$ and $N_{\rm{RF}}$ represent the numbers of antennas and RF chains at the BS or the UE, respectively.
% $O(GKN^2_{\rm{RF}}N_{\rm{s}}(N+N_{\rm{s}}))$

\subsection{Network Training}

We adopt the end-to-end training strategy to train the proposed pilot network, VQ-VAE, HB-GNN, and HC-GNN jointly.
The loss function is defined as
\begin{align}\label{equ:loss_func}
    \mathcal{L} = \alpha\mathcal{L}_{\mathrm{V}} - R
\end{align}
where $\alpha$ is the weight factor keeping fixed during the training phase and the optimal value of $\alpha$ can be obtained by empirical results. The first term of (\ref{equ:loss_func}) corresponds to supervised learning for implicit CSI transmission, which ensures that the received pilot signals and the codewords in the VQ-VAE have similar distributions. The second term of (\ref{equ:loss_func}) pertains to unsupervised learning for maximizing the transmission rate.
In the training phase, we only need to collect channel samples in a targeted environment which serve as the input to the proposed neural network, without the need of creating labels. 
Note that the proposed DNN is site-specific and needs to be retrained if the channel statistics vary. In practice, the channel statistics usually evolve slowly and remain almost unchanged in a long period, indicating that there is no need to execute the retraining operation frequently.

%The proposed method is first trained offline then put into the deployment phase. \textcolor{red}{In addition, We remark that the whole network is site-specific and needs to be retrained if the environment statistics change. In practice, the channel environment usually evolves slowly and remains almost static in a long period, indicating that there is no need to execute the retraining operation frequently.} 

%In the training phase, a dataset $\mathcal{H}$ which contains a large amount of channel sets $\{\mathbf{H}[k]\}_{m=1}^M$ for $M$ subcarriers is adopted. To solve the optimization problem in (\ref{equ:problem}), we propose to train our DNN in an end-to-end strategy. Specifically, the pilot set, VQ-VAE and GNNs are trained jointly in an unsupervised manner to maximize the objective function, the loss function is facilitated to be the sum of $-\mathbb{E}[\hat{R}[1],\hat{R}[2],\dots,\hat{R}[M]]$ and the MSE loss of VQ-VAE, where $\hat{R}[k]$ is the spectral efficiency for the $m$-th subcarrier under the predicted precoder and combiner.

%%%%%%%%%%%%%%%%%%% table %%%%%%%%%%%%%%%%%%%%%%%%%%%
\begin{table}[t]
\newcommand{\tabincell}[2]{\begin{tabular}{@{}#1@{}}#2\end{tabular}}
\caption{Parameter configurations of VQ-VAE}\label{VQVAE_parameter}
\vspace{-0.4cm}
\begin{center}
\begin{tabular}{c|c c c c c c c c c}
\hline
$B$ & 32 & 64 & 96 & 128 & 192 & 256 & 512 & 768 & 1024\\
\hline
$D$ & 2 & 4 & 8 & 4 & 8 & 16 & 16 & 8 &  16\\
\hline
$V$ & 32 &32& 32 & 16 & 16 & 16 & 8 & 4 & 4\\
\hline
\end{tabular}\\
\begin{flushleft}
Note: Since the feedback vector $\mathbf{q}$ is actually a set of indices of codewords and $\mathcal{R}$ consists of the real and imaginary components, the parameters satisfy the equation $B = \frac{2 K_p N_{\rm{s}}L}{V} \log_{2}D $.
\end{flushleft}
\label{tab1}
\end{center}
\vspace{-0.8cm}
\end{table}
%%%%%%%%%%%%%%%%%%% table %%%%%%%%%%%%%%%%%%%%%%%%%%%
%%%%%%%%%%%%%%%%%%% table %%%%%%%%%%%%%%%%%%%%%%%%%%%

\begin{table}[t]
\newcommand{\tabincell}[2]{\begin{tabular}{@{}#1@{}}#2\end{tabular}}
\caption{Complexity comparison}\label{complexity}
\vspace{-0.8cm}
\begin{center}
\begin{tabular}{c|c}
\hline
Hybrid beamforming technique & Complexity\\
\hline
HB-GNN / HC-GNN & $O(G N^2_{\rm{RF}} (K  N^2_{\rm{s}}+ N_{\rm{s}}N+N^2 ))$ \\
MLP & $O(G N^2_{\rm{RF}} (K^2  N^2_{\rm{s}}+ KN_{\rm{s}}N+N^2 ))$\\
MO & $O(IK^2N_{\rm{RF}}N^2_{\rm{s}}N^3)$ \\
Method proposed in \cite{c3}  & $O(D K N^2 \bar{M}^2 \bar{C}^2)$\\
\hline
\end{tabular}\\
\begin{flushleft}
Note: $N$ and $N_{\rm{RF}}$ represent the numbers of antennas and RF chains at the BS or the UE, respectively. $I$ represents the number of iterations.
% and usually ranges from $50$ to $500$ in the simulation
$D$ is the number of convolutional layers. $\bar{M}$ and $\bar{C}$ represent the average kernel size and the number of channels in the convolutional network, respectively.
\end{flushleft}
\label{tab2}
\end{center}
\vspace{-0.8cm}
\end{table}
%%%%%%%%%%%%%%%%%%% table %%%%%%%%%%%%%%%%%%%%%%%%%%%

\section{Numerical Results}
%是单做一个场景 I3 ，还是做两个场景 O1
\subsection{Dataset Description and Simulation Settings}
We perform extensive simulations based on the public datasets of DeepMIMO I3 \cite{deepmimo}. The data associated with the BS \#2 is adopted in the experiment. The downlink carrier frequency and the bandwidth are the $60$ GHz band \footnote{The proposed approach is applicable to a wide range of frequency bands including the centimeter-wave and millimeter-wave bands.} and $100$ MHz, respectively.
We set $N_{\rm{t}} = 64$, $N_{\rm{RF},\rm{t}} = 4$ for the BS and $N_{\rm{r}} = 4$, $N_{\rm{RF},\rm{r}} = 2$ for the UE. The number of subchannels $K$ is $128$ and there are $N_{\rm{s}}=2$ data streams. The noise power spectral density (PSD) and pilot transmission power are set to be -161 dBm/Hz and 10 dBm, respectively, if not specified otherwise. For pilot transmission, we only use $K_p = 16$ subchannels with $M=8$, thus $\mathcal{K}_{p}= \{1,9,\dots,121\}$. The pilot length is set to be $L=16$. The parameter configurations of the proposed VQ-VAE for different feedback overhead $B$ are shown in Table~\ref{VQVAE_parameter}, where the codebook size and the codeword length are denoted as $D$ and $V$, respectively.
Based on simulation trials, the number of layers of aggregation and combination is set to be $G=4$ for both HB-GNN and HC-GNN, and the weight factor $\alpha$ is set to be $0.2$.
We use 60\% samples for training, 20\% for validation, and 20\% for testing.

\begin{figure}[t]
\begin{centering}
\includegraphics[width=0.4 \textwidth]{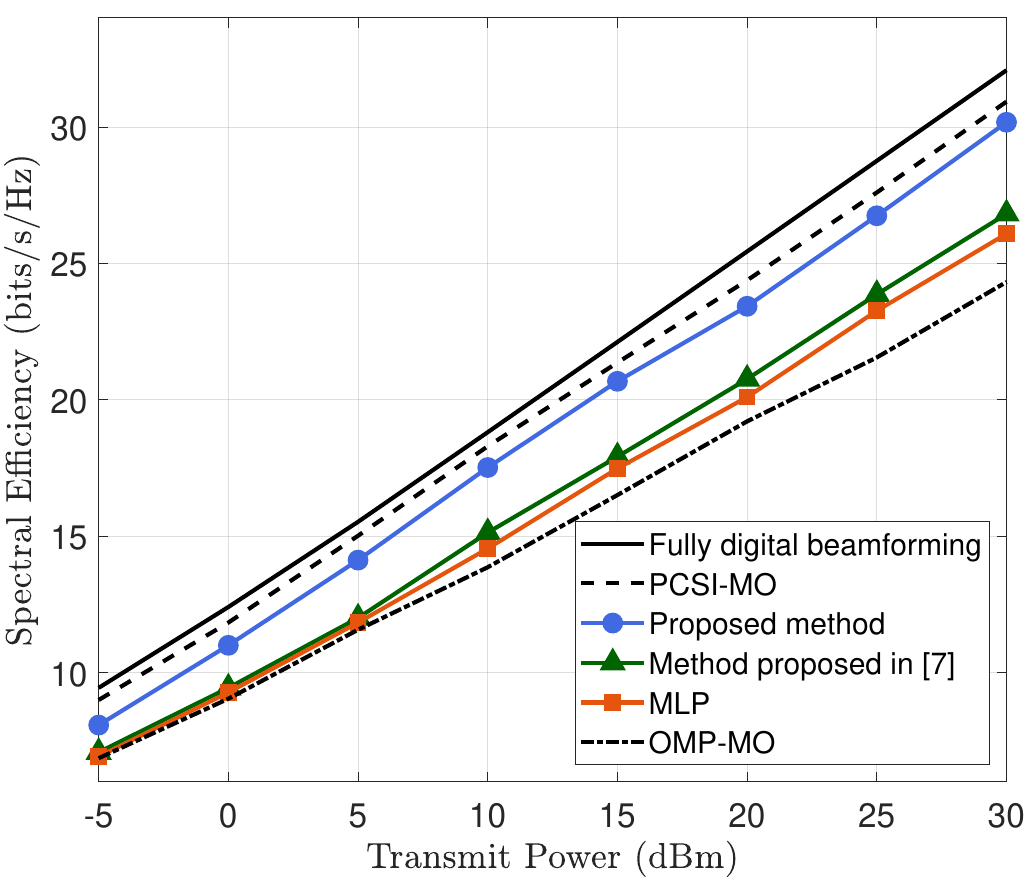}
\vspace{-0.4cm}
\caption{Spectral efficiency v.s. transmit power}\label{fig:SNR}
\end{centering}
\vspace{-0.4cm}
\end{figure}

\begin{figure}[t]
\begin{centering}
\includegraphics[width=0.4 \textwidth]{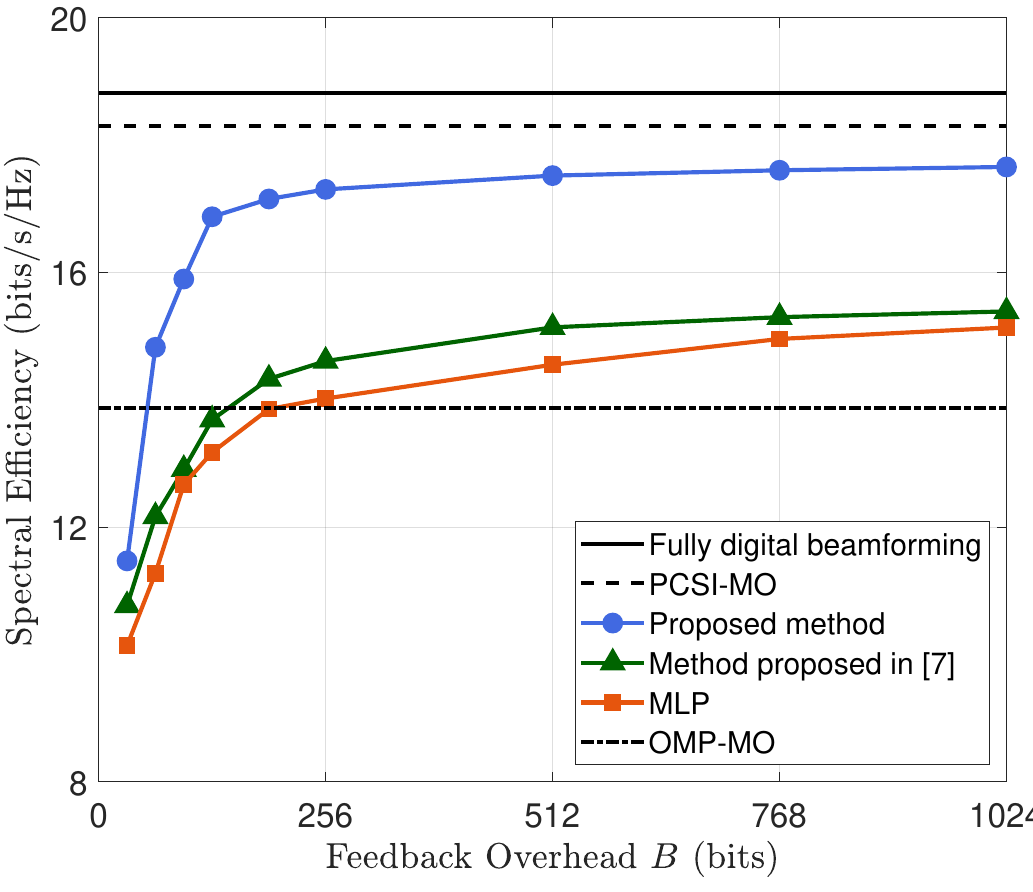}
\vspace{-0.4cm}
\caption{Spectral efficiency v.s. feedback overhead}\label{fig:BIT}
\end{centering}
\vspace{-0.6cm}
\end{figure}

\subsection{Performance Evaluation of the Proposed Method}
To verify the effectiveness of our proposed method, several benchmarks are selected for comparison, including manifold optimization (MO) with perfect CSI (PCSI), MO with orthogonal matching pursuit (OMP)-based channel estimation, DL method proposed in \cite{c3}, and the MLP method whose structure has been utilized in many existing works\cite{c1,reviewer1_recommended3,reviewer_joint}.
The fully digital beamforming with PCSI is also considered as a performance upper bound. 
Here, the MLP method consists of the same structure as the proposed method, but the initialization layer and $G$ aggregation and combination layers in the beamforming network are replaced by $G+1$ fully-connected layers with an activation function. The DL architecture proposed in \cite{c3} primarily comprises convolutional layers and assumes the availability of PCSI at the receiver.
An overview of the complexities of different beamforming schemes is presented in Table~\ref{complexity}.
%and the computational complexity is about $O(G N^2_{\rm{RF}} (K^2  N^2_{\rm{s}}+ KN_{\rm{s}}N+N^2 ))$.
% $O(GK^2N^2_{\rm{RF}}N^2_{\rm{s}})$

%\textcolor{red}{We also evaluate a state-of-the-art DL-based multilayer perceptron (MLP) method [xx], which proposes to realize the auto-encoder and predict the hybrid beamformer by linear layers.}

Fig.~\ref{fig:SNR} shows the average spectral efficiency of proposed method and benchmarks with respect to the transmit power. In the simulation, the feedback overhead is fixed to $512$ bits for both the proposed method and MLP method, whereas the other schemes consider infinite-capacity feedback links. It is observed that our method can significantly outperform other methods that require channel estimation, and achieve performance close to the fully digital beamforming (upper bound). The small performance gap between our method and PCSI-MO may come from the imperfect CSI caused by the limited pilot length, noise to pilot signals, and quantization error in the feedback.

% In particular, the average spectral efficiencies of our method are 3.0 bit/s/Hz and 3.3 bit/s/Hz higher than that of the MLP framework when the transmit power are 10 dBm and 20 dBm, respectively.
%关于和MLP的对比

Next, we evaluate the performance of the DL-based methods with different feedback overheads in Fig. \ref{fig:BIT}, where the transmit power is fixed to 10 dBm.  
It is seen that 256 bits per pilot-bearing subchannel are already sufficient for the proposed method to achieve satisfactory performance. We also observe that the proposed method can outperform MO with OMP-based channel estimation with a feedback overhead of only 64 bits per pilot-bearing subchannel. 
% In conclusion, our method can consistently outperform all other methods that perform explicit channel estimation and can closely approach the performance upper bound with low feedback overhead.

% In addition, when the feedback overhead $B$ is lager than $32$ bits, 
%Moreover, we can also observe that there is still a performance gaps between the proposed GNN architecture with VQ-VAE and the MLP framework \textcolor{red}{in different feedback overhead.}

%这句话放到introduction！！！！
% Here, all mentioned conventional methods have to obtain the explicit channel matrix before beamforming design, 

\section{Conclusion}
In this paper, we investigate the joint design of pilot, CSI feedback, and hybrid beamforming for the FDD MIMO-OFDM system.
A novel DL-based method is proposed, which consists of a PN, an FN, and an HBC-GNN. Therein, the PN uses learned pilot for better CSI acquisition, and the FN via VQ-VAE is designed to improve the feedback efficiency in the limited feedback scenario. Then the HBC-GNN outputs the hybrid beamformer and combiner based on the processed signals at the PN and FN. Simulation results demonstrate the superior performance of our method compared with representative conventional counterparts.
\vspace{-0.0cm}

\bibliographystyle{IEEEtran}
\bibliography{refer}
	
\end{document}